\documentclass{article}

\pdfoutput=1
\usepackage{PRIMEarxiv}

\usepackage[utf8]{inputenc} 
\usepackage[T1]{fontenc}    
\usepackage{hyperref}       
\usepackage{url}            
\usepackage{booktabs}       
\usepackage{amsfonts}       
\usepackage{nicefrac}       
\usepackage{microtype}      
\usepackage{lipsum}
\usepackage{fancyhdr}       
\usepackage{graphicx}       
\graphicspath{{media/}}     
\usepackage{natbib}
\setcitestyle{authoryear} 
\usepackage[figuresright]{rotating}
\usepackage{amsmath}
\usepackage{amsfonts}

\usepackage[T1]{fontenc}
\usepackage{newpxmath}
\usepackage{hyperref}
\hypersetup{
    colorlinks,
    linkcolor={blue},
    citecolor={blue},
    urlcolor={black}
}

\usepackage{lscape}
\usepackage{longtable}
\usepackage{multirow}
\usepackage{threeparttable}
  
\title{Inverse Probability Weighting for Recurrent Event Models}

\author{
  Jiren Sun \\
  Department of Biostatistics and Medical Informatics \\
  University of Wisconsin--Madison \\
  Madison, WI \\
\texttt{jiren.sun@wisc.edu} \\
   \And
  Tobias M{\"u}tze \\
  Statistical Methodology \\
  Novartis Pharma AG \\
  Basel, Switzerland \\
  \texttt{tobias.muetze@novartis.com} \\
     \And
  Richard Cook \\
  Department of Statistics and Actuarial Science \\
  University of Waterloo \\
  Waterloo, Canada \\
    \texttt{rjcook@uwaterloo.ca} \\
   \And
  Tianmeng Lyu \\
  Statistical Methodology \\
  Novartis Pharmaceuticals Corporation \\
  East Hanover, NJ \\
    \texttt{tianmeng.lyu@novartis.com} \\
}

\begin{document}
\maketitle

\begin{abstract}
Recurrent events are common and important clinical trial endpoints in many disease areas, e.g., cardiovascular hospitalizations in heart failure, relapses in multiple sclerosis, or exacerbations in asthma. 
During a trial, patients may experience intercurrent events, that is, events after treatment assignment which affect the interpretation or existence of the outcome of interest. 
In many settings, a treatment effect in the scenario in which the intercurrent event would not occur is of clinical interest. 
A proper estimation method of such a hypothetical treatment effect has to account for all confounders of the recurrent event process and the intercurrent event. 
In this paper, we propose estimators targeting hypothetical estimands in recurrent events with proper adjustments of baseline and internal time-varying covariates.
Specifically, we apply inverse probability weighting (IPW) to the commonly used Lin-Wei-Yang-Ying (LWYY) and negative binomial (NB) models in recurrent event analysis. Simulation studies demonstrate that our approach outperforms alternative analytical methods in terms of bias and power.
\end{abstract}


\section{Introduction}

Recurrent events are instances where a particular event may recur within the same subject. Medical examples include hospitalizations due to heart failure, relapses in multiple sclerosis, or exacerbations in asthma  \citep{rogers2014analysing, cohen2011multiple, keene2007analysis}. Between-subject heterogeneity at risk for recurrent events results in a within-subject dependence in the event times, which must be addressed in analyses to mitigate under-estimation of standard errors, artificially narrowed confidence intervals, and inflated Type I error rates \citep{amorim2015modelling}. The two most popular methods for analyzing recurrent event data in clinical trials are the Lin-Wei-Yang-Ying (LWYY) model and the negative binomial (NB) model \citep{lin2000semiparametric,lawless1987negative}. Both models assume multiplicative treatment effects defined as the ratio of the marginal event rates between the two arms. The semiparametric LWYY model uses a partial likelihood score function corresponding to that of the Andersen-Gill model \citep{andersen1982cox}. Inference in the LWYY model is based on a robust sandwich variance estimate to account for the association of the recurrent events within subjects. In contrast, the NB model is a Poisson–gamma mixed effect model, which assumes the number of events within a subject follows a Poisson distribution conditional on a Gamma-distributed frailty---this means that given the random effect, events are treated as arising from a Poisson process. Integrating over the frailty yields a marginal distribution for event counts that follows the negative binomial distribution.

The International Council of Harmonization (ICH) E9(R1) addendum proposes a framework for aligning the estimand (i.e., the target of estimation reflecting the clinical question of interest) with the trial design, conduct, and analysis \citep{international2019e9}. The addendum emphasizes the importance of identifying the intercurrent events and reflecting them in the estimand. Intercurrent events are events that occur after treatment assignment and ``affect either the interpretation or the existence of the measurements associated with the clinical question of interest.'' Examples of intercurrent events include treatment discontinuation, treatment switching, rescue medication, death, etc. For an in-depth discussion of ICH E9(R1) and its role in clinical trials, we refer to existing literature \citep{akacha2017estimands, clark2022estimands, keene2023estimands}. One of the clinical questions of interest discussed in ICH E9(R1) is the so-called hypothetical estimand, e.g., ``What is the treatment effect had subjects not experienced the intercurrent event?'' This type of question is commonly answered using causal inference methodology (e.g., inverse probability weighting, g-computation, etc) or missing data methodology (e.g., multiple imputation) \citep{olarte2023hypothetical,lasch2023simulation}.

Both the recurrent event process and the intercurrent events process may be influenced by treatment assignment, baseline covariates, and internal time-varying covariates. Failing to properly account for these covariates can result in biased estimates of the hypothetical estimand of interest. While treatment assignment and baseline covariates can be directly incorporated into the regression model for recurrent events, internal time-varying covariates, which may be affected by treatment, cannot be directly adjusted for in the model. To address this challenge, in this paper, we draw from the principles of marginal structural models \citep{hernan2000marginal, robins2000marginal} and apply inverse probability weighting (IPW) to the widely used LWYY and NB models to address hypothetical questions in clinical trials. This approach involves artificially censoring subjects at their intercurrent event time, retaining only events before the intercurrent event, and assigning (time-dependent) weights to each subject to create a pseudo-population free of intercurrent events and independent of measured confounders. In this pseudo-population, we can conduct the conventional regression analysis, regressing the outcome against the treatment without including internal time-varying covariates. In the presence of the intercurrent event, the IPW approach has been previously employed to address the hypothetical question for survival outcomes \citep{latimer2016treatment, latimer2018assessing} and continuous outcomes \citep{olarte2023hypothetical,lasch2023simulation}, but to the best of our knowledge, its application to the recurrent event setting, especially concerning the LWYY and NB models, has not been studied yet. Motivated by an example described in Section \ref{sec:example:ms}, we discuss the methods under a setting of a two-arm clinical trial with treatment switching as the intercurrent event, but the proposed methods can also be applied to target hypothetical estimands with other intercurrent events.

The remainder of this paper is structured as follows. In Section \ref{sec:example}, we describe two motivating examples. In Section \ref{sec:review}, we illustrate how the LWYY and NB models can be derived from a unified framework. Section \ref{sec:ipw} describes the proposed approach to apply IPW to the LWYY and NB models, building upon the framework outlined in Section \ref{sec:review}. A simulation study is presented in Section \ref{sec:simulation}. For illustration purposes, the application of the proposed methods to analyze a synthetic clinical trial dataset is presented in Section \ref{sec:asthmaanalysis}, and our concluding remarks are included in Section \ref{sec:conclusion}.

\section{Motivating example}
\label{sec:example}
\subsection{Clinical trial in multiple sclerosis}
\label{sec:example:ms}
Multiple sclerosis (MS) is a disease where the body's immune system attacks the protective covering of the nerves, causing problems like muscle weakness, tiredness, and difficulty with coordination. The EXPAND study assessed the effect of Siponimod in treating secondary progressive multiple sclerosis \citep{kappos2018siponimod}. The recurrent relapses in multiple sclerosis were among the secondary outcomes. During the double-blind treatment period, subjects were provided with the option to switch to Siponimod or to stop study treatment and either remain untreated or receive another disease-modifying therapy. It is worth clarifying that, we use the term treatment switching to avoid confusion with trials employing a crossover design, where each subject receives both treatments, acting as their own control.

\subsection{Clinical trial in asthma}
\label{sec:example:asthma}

A second motivating example is a Phase III asthma trial. In this example, 612 patients were assigned to drug X, while 606 patients were assigned to the control group. All participants were expected to be monitored for 26 weeks, although they could discontinue treatment or withdraw from the trial at any time. The primary endpoint is the total number of asthma exacerbations (across all severity levels) up to Week 26. For illustration purposes, we focus on the intercurrent event of treatment discontinuation. In Section \ref{sec:asthmaanalysis}, we illustrate the methods described in Section \ref{sec:ipw} using synthetic data from this clinical trial. 

\section{Reviews of statistical models for recurrent events}
\label{sec:review}
In this section, we present Poisson, NB, and LWYY models from a unified framework without taking into account intercurrent events. 

\subsection{A unified framework}
We let $\{N(t), t \geq 0 \}$ be a right-continuous counting process representing the number of events in $[0,t]$. Let $\Delta N(t)=N(t+\Delta t^{-})-N(t^{-})$ denote the number of events in $[t,t+\Delta t)$. Let $H(t)=\{N(s): 0 \leq s <t ; X(s): 0 \leq s \leq t \}$ denote the history of the process by time $t$, where $X(s)$ includes fixed and time-varying covariates related to the recurrent event occurrences. 
Let $$\lambda\{t \mid  H(t)\}=\lim_{\Delta t \rightarrow 0} \frac{P\{\Delta N(t)=1 \mid  H(t)\}}{\Delta t}$$ be the event intensity function, which is the instantaneous probability of an event occurring at time $t$, conditional on the process history. The intensity function defines an event process. 
Finally, let the at-risk indicator be $Y(t)=I(\hbox{process is observed at time } t)$.
If the intensity function is specified in terms of a parameter $\theta$, using the counting process notation, the log-likelihood can be expressed as 
\begin{eqnarray}
    l(\theta)=\int_{0}^{\infty} Y(t)\hbox{log} \lambda\{t \mid  H(t);\theta\}dN(t) - \int_{0}^{\infty} Y(t) \lambda\{t \mid  H(t);\theta\}dt \label{logllh}.
\end{eqnarray}
Details of the log-likelihood derivation are provided in Section A of the Supplementary Materials.

Let $\mu(t)=E\{N(t)\}$ denote the mean function and let $\rho(t)=d\mu(t)/dt$ be the rate function. Treatment effects are typically evaluated based on marginal features of the process and are usually assumed to be multiplicative, which gives the model form $\rho_X(t) = \hbox{exp}(X^{T} \beta) \rho_{0}(t)$ where $\rho_{0}(t)$ is the baseline event rate and $X$ denotes treatment assignment and baseline covariates. We are primarily interested in estimating $\beta$, where $\exp(\beta)$ represents the multiplicative effect on the event rate, conditional on $X$. Internal time-varying covariates will be addressed in Section \ref{ipcwsection}.

The rate function $\rho(t)$ is the expectation of the intensity function $\lambda\{t \mid  H(t)\}$ over the event history $H(t)$. However, $\rho(t)$ does not fully specify the likelihood. There are two common approaches to establishing a connection between $\rho(t)$ and $\lambda\{t \mid  H(t)\}$. The first approach is to assume $\lambda\{t \mid  H(t)\}=\rho(t)$, which implies that the probability of an event in $(t,t+\Delta t)$ may depend on $t$ but is independent of $H(t)$. When events are defined under this assumption, it is referred to as a Poisson process. Poisson models for recurrent events are discussed in Section \ref{sec:poisson}. The second approach involves assuming that $\lambda\{t \mid  H(t), \gamma\} = \gamma  \rho(t)$, where $\gamma$ represents a random effect specific to each subject. This assumption implies that, given the value of $\gamma$, the event process follows a Poisson distribution with a rate of $\gamma  \rho(t)$. In other words, the information contained in $H(t)$ is assumed to be captured by $\gamma$. The negative binomial model is an example of the second approach by assuming that $\gamma$ follows a gamma distribution with a mean of 1 and a variance of $\phi$, as presented in Section \ref{sec:nb}.

\subsection{Poisson models}\label{sec:poisson}
Suppose there are subjects indexed by $i = 1, \dots, m$, each characterized by the covariate vector $X_{i}$, follow-up time $\tau_i$, and total number of recurrent events $n_i$. By assuming $\lambda\{t \mid  H(t);\theta\}=\rho(t;\theta)$, according to formula (\ref{logllh}), the log-likelihood $l_{P}(\theta)$ is 
$$l_{P}(\theta)  = \sum_{i=1}^{m} \int_{0}^{\tau}Y_{i}(t)  \{ \hbox{log} \rho_{i}(t;\theta)dN_{i}(t)- \rho_{i}(t;\theta)dt \} $$
where $Y_{i}(t)=I(t \leq \tau_i)$ and $\tau=\hbox{max}\{\tau_1, \dots, \tau_m\}$.

\subsubsection{Parametric Poisson regression}
We let $\theta=(\alpha^T,\beta^T)^T$ denote the full vector of parameters, where the baseline rate function is parameterized by $\alpha$, and $\beta$ is a vector of regression coefficients. Based on the likelihood, the estimating equations for $\alpha$ and $\beta$ are 
\begin{eqnarray}
    U^{(P)}_{\alpha}(\theta)&=&\frac{\partial l_{P}(\theta)}{\partial \alpha}=\sum_{i=1}^{m}\int_{0}^{\tau} Y_{i}(t) \frac{\partial \hbox{log}\rho_{0}(t;\alpha) }{\partial \alpha} \{ dN_{i}(t) - \rho_{i}(t;\theta)dt \}=0, \nonumber \\
    U^{(P)}_{\beta}(\theta)&=&\frac{\partial l_{P}(\theta)}{\partial \beta}=\sum_{i=1}^{m}\int_{0}^{\tau} Y_{i}(t)  X_{i}  \{ dN_{i}(t) - \rho_{i}(t;\theta)dt \}=0 .\label{betascore}
\end{eqnarray}

Parametric Poisson regression can be implemented in R using \texttt{glm()}. Many software packages, including \texttt{glm()}, by default assume $\rho_{0}(t;\alpha)=\alpha$, which means the model only counts the total number of events each subject $i$ experienced during the entire follow-up $\tau_i$, but ignores when those events occurred.

\subsubsection{Andersen-Gill model: A semiparametric Poisson model}
In parametric Poisson regression, $\rho_{0}(t)$ is assumed to be parametric and is usually constant. The baseline rate function, however, can also be left unspecified. The Anderson-Gill model is one such Poisson model where the baseline rate is unspecified \citep{andersen1982cox}. It is an extension of the Cox model to recurrent events, with the corresponding estimating equation (profile score) given as follows:
\begin{eqnarray}
    U^{(A)}_{\beta}(\theta) = \sum_{i=1}^{m}  \int_{0}^{\tau} Y_{i}(t) \left\{X_{i} -  \frac{\sum_{i=1}^{m} Y_{i}(t)X_{i} \hbox{exp}(X^{T}_{i} \beta)}{\sum_{i=1}^{m} Y_{i}(t) \hbox{exp}(X^{T}_{i} \beta)}  \right\} dN_{i}(t) =0 \label{lwyybetascore}.
\end{eqnarray}
The derivations of the estimating equations are provided in Section B of the Supplementary Materials.

The Anderson-Gill model can be fitted in R using \texttt{coxph(Surv(start, stop, status)\allowbreak $\sim$\allowbreak covariates, \allowbreak data)}, where \texttt{start} and \texttt{stop} define the time interval boundaries for each recurrent event segment, and \texttt{status} is the event indicator (1 if the recurrent event occurred in the interval, 0 if censored).

The Anderson-Gill model assumes that any effect of previous recurrent events on future occurrences is fully explained by the observed covariates $X$. With $X$ being time invariant, this corresponds to the independent increments structure of the Poisson process. However, this assumption may be unrealistic, as the observed covariates might not fully account for the dependence among recurrent events. To address this issue, one strategy is to use robust inference techniques that can account for within-subject dependence \citep{lin2000semiparametric}.

\subsubsection{LWYY model}
The LWYY model \citep{lin2000semiparametric} models the marginal rate of events: $\rho_X(t) = \hbox{exp}(X^{T} \beta) \rho_{0}(t)$. It adopts the partial-likelihood score function (\ref{lwyybetascore}) for the Andersen-Gill model as an estimating function and uses a robust sandwich estimator for variance estimation. The robust inference procedure allows for arbitrary correlation structure between the recurrent events, which relaxes the assumption of the Anderson-Gill model that the future occurrence of recurrent events depends on the past only through the observed covariates $X$.

The LWYY model can be fitted in R using \texttt{coxph(Surv(start, stop, status)\allowbreak $\sim$\allowbreak covariates + \allowbreak cluster(id), \allowbreak data)\allowbreak}, where \texttt{id} represents the subject identifier.

\subsection{Negative binomial (NB) models}\label{sec:nb}
Another way to account for within-subject correlation is to introduce a subject-specific frailty term to the intensity function so as to induce correlations between the recurrent events. Specifically, we assume $\lambda\{t \mid  H(t), \gamma\}=\gamma  \rho(t) =\gamma  \hbox{exp}(X^{T} \beta) \rho_{0}(t)$, where $\gamma$ is an non-negative random variable following a parametric distribution. In the NB models, $\gamma$ follows a gamma distribution with a mean of 1 and a variance of $\phi$. 

In conventional practice, when fitting a negative binomial model, the standard software typically sets $\rho_{0}(t)=\alpha$ (e.g., using \texttt{glm.nb()} in R). The estimating equations for $\phi$, $\alpha$, and $\beta$ based on the likelihood are
\begin{eqnarray}
    U^{NB}_{\phi}(\theta, \phi) &=& \sum_{i=1}^{m} \left[ \phi^{-2}\hbox{log}\left\{ 1+\phi\mu_{i}(\tau_i) \right\} - (n_i+\phi^{-1}) \frac{\mu_{i}(\tau_i)}{1+\phi \mu_{i}(\tau_i)} + \sum_{j=0}^{n_{i}^{*}} \frac{j}{1+\phi j} \right] =0 ,\nonumber \\ 
    U^{NB}_{\alpha}(\theta, \phi) &=&  \sum_{i=1}^{m} \left[  \left\{ \sum_{j=1}^{n_i} \frac{\partial \rho_{0}(t_{ij}) / \partial \alpha  }{\rho_{0}(t_{ij})}  \right\} - \frac{(1+ \phi n_i ) \hbox{exp}(X^{T}_{i} \beta ) }{1+ \phi\mu_{i}(\tau_i)  } \frac{\partial \mu_{0}(\tau_i)}{\partial \alpha}  \right] = 0 ,\nonumber \\ 
    U^{NB}_{\beta}(\theta, \phi) &=& \sum_{i=1}^{m} \frac{n_i - \mu_{i}(\tau_i) }{1+ \phi \mu_{i}(\tau_i) }  X_i = 0.\label{nb1betascore}
\end{eqnarray}
where $n_{i}^{*}=\hbox{max}\{0, n_i-1\}$.

A detailed derivation can be found in Section C of the Supplementary Materials. The R function \texttt{glm.nb()} is commonly used to fit the NB model with a constant baseline.

Choosing a constant baseline $\rho_{0}(t) = \alpha$ implies that the model accounts only for the total number of events experienced by each subject over the entire follow-up period, without considering the timing of these events. However, the timing information can be incorporated by leaving $\rho_{0}(t)$ unspecified and estimating it semiparametrically. A pseudo-log-likelihood of the NB model with an unspecified baseline is given by \cite{cook2007statistical}:
\begin{eqnarray}
    l_{NB}(\beta;\phi)= \sum_{i=1}^{m} \int_{0}^{\infty} Y(t)\hbox{log} \lambda_i\{t \mid  H^{*}_i(t);\beta,\phi\}dN(t) - \int_{0}^{\infty} Y(t) \lambda_i\{t \mid  H^{*}_i(t);\beta,\phi\}dt \label{nb2esteqn}
\end{eqnarray}
where the intensity function is given by $\lambda_{i}\{t \mid  H^{*}_{i}(t);\beta,\phi \}=\left\{ \frac{1+ \phi N_{i}(t^{-}) }{1+ \phi \mu_{i}(t) }  \right\}  \hbox{exp}(X^{T}_{i} \beta) \rho_0(t)$, and the event history $H^{*}_i(t)$ involve the total number of events up to but not including $t$, the corresponding event times, and the baseline covariates $\{N_{i}(t^{-}), t_{i1}, \dots, t_{iN_{i}(t^{-})}, X_i \}$.  The baseline intensity $\rho_0(t)$ can be estimated semiparametrically by $\hat \rho_0(t)=\frac{\sum_{i=1}^{m} Y_{i}(t)  dN_{i}(t)}{\sum_{i=1}^{m} Y_{i}(t) \hbox{exp}(X^{T}_{i} \beta)}$. The detailed derivation can be found in Section D of the Supplementary Materials.

Standard software does not fit the NB model with an unspecified baseline, so custom code is required. However, common optimization tools (e.g., \texttt{optim()} in R) can be used to maximize the pseudo-log-likelihood \eqref{nb2esteqn} with respect to $\beta$ and $\phi$, obtaining the maximum pseudo-likelihood estimate for $\beta$.

\section{Application of IPW}
\label{sec:ipw}
In the presence of treatment switching, we extend LWYY and NB models to answer the hypothetical question: What would the treatment effect be if subjects were not allowed to switch? Both the recurrent event process and treatment switching process may be impacted by treatment assignment, baseline covariates $X$, and internal time-varying covariates $L(t)$, so failing to properly adjust for such confounders will lead to biased estimates for the hypothetical estimand of interest. Treatment assignment and baseline covariates $X$ can be directly adjusted in the regression model for recurrent events. However, because $L(t)$ can be impacted by the treatment, it cannot be directly adjusted for in the regression model, and thus we propose to apply the IPW approach to adjust for such confounders. 

\subsection{Introduction to IPW}\label{ipcwsection}
The Inverse Probability Weighting (IPW) enables us to account for the effects of $L(t)$ without including it as a regressor. In this approach, we artificially censor subjects at the time of switching, discard events occurring after the switching, and assign weights to each subject based on the inverse probability of the subject remaining on their assigned treatment (abbreviated as remaining unswitched henceforth). Because treatment switching is a time-dependent process, the weights need to be time-dependent as well. By using the IPW approach, the follow-up time interval for each subject $i$ is subject to ``truncation'' due to switching. This truncated follow-up interval is represented as $(0, \min(\tau_i, S_i))$, where $S_i$ denotes the switching time for subject $i$, and is set to infinity if subject $i$ does not switch.

The rationale behind the use of inverse weighting can be intuitively explained as follows: A generic subject with covariates $B_i(t)=(X_i^{T}, L_i^{T}(t))^T$ has a probability $P\{S_i \geq t \mid  B_i(t)\}$ of remaining unswitched at time $t$. This implies that, on average, one out of every $1/P\{S_i \geq t \mid  B_i(t)\}$ subjects with covariates $B_i(t)$ remains unswitched. By assigning a weight of $w_{i}(t)=1/P\{S_i \geq t \mid  B_i(t)\}$, where $t \leq \hbox{min}(\tau_i, S_i) $, we construct a pseudo-population in which all subjects remain unswitched.

IPW relies on some assumptions, including no unmeasured confounders and the positivity assumption. The no unmeasured confounders assumption implies that, given $B(t)$, the recurrent event outcomes are independent of the switching process. Positivity assumption entails that for each level of $B(t)$, the probability of remaining unswitched $P(S_i \geq t)$ is greater than zero. Further discussion of IPW assumptions can be found in \cite{cole2008constructing}.

We re-define the at-risk process in the context of treatment switching: $Y^{*}_{i}(t)=I(t \leq \hbox{min}\{\tau_{i}, S_i\})$, where $\tau_i$ is the follow-up time of subject $i$ and $Y^{*}_{i}(t)=1$ indicates that subject $i$ remains on the assigned treatment and remains under observation. Different from $Y_{i}(t)=I(t \leq \tau_i)$, the subject with $S_i < \tau_i$ will remain at risk under $Y_{i}(t)$ for $S_i < t < \tau_i$, but is not at risk under $Y^{*}_{i}(t)$ once he or she switches the treatment. Mathematically, $Y^{*}_{i}(t)=Y_{i}(t)I(t \leq S_i)$. Because $B(t)$ influences both $N(t)$ and $Y^{*}(t)$, $N(t)$ and $Y^{*}(t)$ are not independent. However, under the assumption of no unmeasured confounders, we consider their conditional independence given the values of $B(t)$.

\subsection{Application of IPW to LWYY (LWYY + IPW)}

The application of IPW to LWYY involves incorporating the weight $w_i(t)$ into the estimating equations for both $d\mu_0(t)$ and $\beta$. 

Treating $d\mu_{0}(t)=\rho_{0}(t)dt$ as a parameter, we obtain the modified estimating equation for $d\mu_{0}(t)$, where each summand is weighted by $w_i(t)$. Specifically, we have the following equation:
\begin{eqnarray}
    \sum_{i=1}^{m} Y^{*}_{i}(t)  w_{i}(t) \{ dN_{i}(t) - d\mu(t)  \} 
    = \sum_{i=1}^{m} Y^{*}_{i}(t) w_{i}(t) \{ dN_{i}(t) - \hbox{exp}(X^{T}_{i} \beta)d\mu_{0}(t)  \} = 0. \label{eq:lwyyipwbas}
\end{eqnarray}

From this, we can derive the profile likelihood estimate of $d\mu_{0}(t)$:
\begin{eqnarray}
    d\widehat \mu_{0}(t; \beta) = \frac{\sum_{i=1}^{m} Y^{*}_{i}(t) w_{i}(t) dN_{i}(t)}{\sum_{i=1}^{m} Y^{*}_{i}(t) w_{i}(t) \hbox{exp}(X^{T}_{i} \beta)} .\label{baselinerateipcw}
\end{eqnarray}

Similarly, the estimating equation for $\beta$ (Equation (\ref{betascore})) needs to be modified to incorporate the weight $w_i(t)$:
\begin{eqnarray}
    \sum_{i=1}^{m}\int_{0}^{\tau} Y^{*}_{i}(t) w_{i}(t) X_{i}\{ dN_{i}(t) - \rho_{i}(t;\beta)dt \}=0. \label{betascoreipcw}
\end{eqnarray}

Plugging (\ref{baselinerateipcw}) to (\ref{betascoreipcw}), we have:
\begin{eqnarray*}
   \sum_{i=1}^{m}  \int_{0}^{\tau} Y^{*}_{i}(t)w_{i}(t) \left\{X_{i} -  \frac{\sum_{i=1}^{m} Y^{*}_{i}(t)  w_{i}(t) X_{i} \hbox{exp}(X^{T}_{i} \beta)}{\sum_{i=1}^{m} Y^{*}_{i}(t) w_{i}(t) \hbox{exp}(X^{T}_{i} \beta)}  \right\}  dN_{i}(t) = 0 \label{lwyyipcw}.
\end{eqnarray*}

This approach is a straightforward extension of the marginal structural Cox model to handle recurrent event scenarios \citep{hernan2000marginal}.

The unbiasedness of the modified estimating equations \eqref{eq:lwyyipwbas} and \eqref{betascoreipcw} can be demonstrated using the law of iterated expectations:
\begin{align*}
    \hbox{E}\{ Y^*_i(t) w_i(t) \, dN_i(t) \}
    &= w_i(t) \, \hbox{E}\{ I(t \leq S_i) Y_i(t) \, dN_i(t) \} \\
    &= w_i(t) \, \hbox{E}\left[ \hbox{E}\{ I(t \leq S_i) Y_i(t) \, dN_i(t) \mid B_i(t) \} \right] \\
    &= w_i(t) \, \hbox{E}\left[ \hbox{E}\{ I(t \leq S_i) \mid B_i(t) \} \right] \,\hbox{E} \left[ \hbox{E}\{ Y_i(t) \, dN_i(t) \mid B_i(t) \} \right] \\
    &= \hbox{E}\{ Y_i(t) \, dN_i(t) \}.
\end{align*}

The LWYY + IPW can be fitted in R using \texttt{coxph(Surv(start, stop, status)\allowbreak $\sim$\allowbreak covariates + cluster(id),\allowbreak data,\allowbreak weights)}, where \texttt{weights} correspond to $w_i(t)$.

\subsection{Application of IPW to NB}
\subsubsection{NB + IPW}
The application of IPW to the NB model involves integrating the weight $w_i(t)$ into both the pseudo-log-likelihood function and the intensity function.

The pseudo-log-likelihood for the NB model (Equation (\ref{nb2esteqn})) is modified as follows:
\begin{eqnarray*}
    \sum_{i=1}^{m} \int_{0}^{\infty} Y^{*}_i(t) w_i(t) \hbox{log} \lambda^{*}_i\{t \mid  H^{*}_i(t);\beta\}dN_i(t) - \int_{0}^{\infty} Y^{*}_i(t) w_i(t) \lambda^{*}_i\{t \mid  H^{*}_i(t);\beta\}dt 
\end{eqnarray*}
where $\lambda^{*}_{i}\{t \mid  H^{*}_{i}(t);\beta \}$ is the modified $\lambda_{i}\{t \mid  H^{*}_{i}(t);\beta \}$ by incorporating $w_i(t)$. Specifically, it is given by $\lambda^{*}_{i}\{t \mid  H^{*}_{i}(t) \}=\left\{ \frac{1+ \phi N_{i}(t^{-}) }{1+ \phi \mu_{i}(t) }  \right\}  \hbox{exp}(X^{T}_{i} \beta) \rho_0^*(t)$, where the modified baseline intensity $\rho_0^*(t)$ can be estimated semiparametrically by $\hat \rho_0^*(t)=\frac{\sum_{i=1}^{m} Y^{*}_{i}(t) w_i(t)  dN_{i}(t)}{\sum_{i=1}^{m} Y^{*}_{i}(t) w_i(t)  \hbox{exp}(X^{T}_{i} \beta)}$. The validity of the modified pseudo-log-likelihood can similarly be established using the law of iterated expectations.

Standard software does not fit NB + IPW, so custom code is required. The R function \texttt{optim()} can be used to maximize the modified pseudo-log-likelihood with respect to $\beta$ and $\phi$, obtaining the maximum pseudo-likelihood estimate of $\beta$.

\subsubsection{Na{\"i}ve NB + IPW}
If a quick estimate of $\beta$ from an inverse probability weighted NB model is desired without custom coding, the commonly used R function \texttt{glm.nb()} can potentially be employed by specifying the \texttt{weights} argument with $w_i(t)$. However, \texttt{glm.nb()} estimates $\beta$ by solving Equation \eqref{nb1betascore}, and the structure of this estimating equation has implications for incorporating $w_i(t)$. Specifically, the estimating equation for $\beta$ from the NB model with a constant baseline (Equation \eqref{nb1betascore}) depends only on the total count of events experienced by each subject and does not account for the timing of events. As a consequence, although each subject $i$ may have time-varying weights over the follow-up period, Equation \eqref{nb1betascore} only allows the assignment of a single weight per subject evaluated at their last follow-up time, $\tau_i$. To incorporate $w_i(\tau_i)$ in an unbiased estimating equation, the analysis must be restricted to non-switchers. This leads to the following modified estimating equations for $\phi$, $\alpha$, and $\beta$:
\begin{eqnarray}
    && \sum_{i=1}^{m} w_{i}(\tau_i) I(S_i \geq \tau_i)\left[ \phi^{-2}\hbox{log}\left\{ 1+\phi\mu_{i}(\tau_i) \right\} - (n_i+\phi^{-1}) \frac{\mu_{i}(\tau_i)}{1+\phi \mu_{i}(\tau_i)} + \sum_{j=0}^{n_{i}^{*}} \frac{j}{1+\phi j} \right] =0 ,\nonumber \\ 
    &&\sum_{i=1}^{m} w_{i}(\tau_i) I(S_i \geq \tau_i) \left[  \left\{ \sum_{j=1}^{n_i} \frac{\partial \rho_{0}(t_{ij}) / \partial \alpha  }{\rho_{0}(t_{ij})}  \right\} - \frac{(1+ \phi n_i ) \hbox{exp}(X^{T}_{i} \beta ) }{1+ \phi\mu_{i}(\tau_i)  } \frac{\partial \mu_{0}(\tau_i)}{\partial \alpha} \right] = 0, \nonumber \\
    &&\sum_{i=1}^{m} w_{i}(\tau_i) I(S_i \geq \tau_i) \frac{n_i - \mu_{i}(\tau_i) }{1+ \phi \mu_{i}(\tau_i) } X_i = 0. \label{nb1ipcw}
\end{eqnarray}

Let $U_{i}^{*}=w_{i}(\tau_i) I(S_i \geq \tau_i) \frac{n_i - \mu_{i}(\tau_i) }{1+ \phi \mu_{i}(\tau_i) } X_i$ and $U_{i}=\frac{n_i - \mu_{i}(\tau_i) }{1+ \phi \mu_{i}(\tau_i) } X_i$. We can show that
\begin{eqnarray*}
    \hbox{E}(U_{i}^{*}) &=& w_{i}(\tau_i)  \hbox{E}\{I(S_i \geq \tau_i)  U_i\} = w_{i}(\tau_i)  \hbox{E} [\hbox{E}\{I(S_i \geq \tau_i)  U_i\mid  B_i(t) \} ] \\
    &=& w_{i}(\tau_i)  \hbox{E} [\hbox{E}\{I(S_i \geq \tau_i)\mid  B_i(t)\}  \hbox{E}\{U_i\mid  B_i(t) \} ] \\
    &=& w_{i}(\tau_i)  \hbox{E} [w_{i}^{-1}(\tau_i)  \hbox{E}\{U_i\mid  B_i(t)\} ] = \hbox{E} [\hbox{E}\{U_i\mid  B_i(t)\} ] = \hbox{E}(U_i)    .
\end{eqnarray*}

Therefore, the modified estimating equation is unbiased for estimating $\beta$; the unbiasedness for estimating $\phi$ and $\alpha$ can be demonstrated using analogous reasoning.

The formulation of the modified estimating equations \eqref{nb1ipcw} involves two steps. First, the analysis is restricted to non-switchers, defined as the set $\{i: S_i \geq \tau_i\}$. Second, for each non-switcher, only the weight $w_{i}(\tau_i)$ is applied. We refer to this approach as the ``Na{\"i}ve NB + IPW.''

\subsection{Comparison of LWYY + IPW, NB + IPW, and Na{\"i}ve NB + IPW}

The three approaches differ in the following ways: First, while NB + IPW and LWYY + IPW include all subjects in the estimating equations, Na{\"i}ve NB + IPW restricts to non-switchers only. Second, the use of weights in the estimating equations differs among the approaches: for each subject $i$, Na{\"i}ve NB + IPW applies a single weight $w_{i}(\tau_i)$; LWYY + IPW uses $w_{i}(t)$ specifically at recurrent event time points; and NB + IPW incorporates all $w_{i}(t)$ over the interval $(0,\min(\tau_i, S_i))$. These differences have important implications for weight estimation, which will be discussed in more detail in the next subsection.

\subsection{Estimation of weights}\label{weightest}

The probability $P\{S_i \geq t \mid  B_i(t)\}$ can be estimated, for instance, through a pooled logistic regression model. For practical convenience, we let the underlying value of $L(t)$ update on a discrete time scale (e.g., weekly). For each subject $i$, their truncated follow-up time interval $(0, \hbox{min}(\tau_i, S_i) )$ is segmented at the time points when the underlying value of $L(t)$ is updated. Each segmented interval of subject $i$ corresponds to a pair of values $(e_{i}(t), B_{i}(t))$, where $e_{i}(t)$ is a binary indicator, which is 1 if the subject $i$ switches the treatment at time $t$, and 0 otherwise. The estimation of weights involves all pairs of $(e_{i}(t), B_{i}(t))$ across segmented intervals from all subjects. It is important to note that, in practice, measurements of $L(t)$ are not conducted as frequently as the underlying value of $L(t)$ is updated. For example, $L(t)$ may be supposed to update each week, while the measurement of $L(t)$ is conducted every three months. For missing values between measurements, they are typically imputed using the last observed value. The truncated follow-up time of each subject is segmented at time points when the underlying value of $L(t)$ is updated, rather than only at time points when the measurement is conducted. 

The pooled logistic regression approach estimates $P\{S_i \geq t \mid  B_i(t)\}$ through the following steps:
\begin{enumerate}
    \item For each subject $i$, gather all pairs $(e_{i}(t), B_{i}(t))$ where $t \leq \min(\tau_i, S_i)$ into $\mathcal{V}_i$, where $\tau_i$ represents the follow-up time. Then combine all $\mathcal{V}_i$ into a single set $\mathcal{V}$ for $i=1, \dots, m$.
    \item Express the pooled logistic regression model as $$\hbox{log}\left[ \frac{ P\{e_{i}(t)=1 \mid  e_{i}(t^{-})=0\} }{1-P\{e_{i}(t)=1 \mid  e_{i}(t^{-})=0\}} \right] = B_{i}^{T}(t)\beta_{p}  $$
    where $(e_{i}(t), B_{i}(t)) \in \mathcal{V}$, and $P\{e_{i}(t)=1 \mid  e_{i}(t^{-})=0\}$ is the conditional probability of observing a switching by time $t$ given that the subject $i$ has not yet switched by $t^{-}$.
    \item To estimate $P\{S_i \geq t \mid  B_i(t)\}$, substitute $B_{i}(s)$ into the model fitted in step (2) and obtain the estimated probability of $e_{i}(s)=0$, given that $e_{i}(s^{-})=0$. Then $\hat P\{S_i \geq t \mid  B_i(t)\}$ can be expressed as the product of the individual probabilities for each $s < t$:  $\hat P\{S_i \geq t \mid  B_i(t)\} = \prod_{s < t}  \hat P\{S_i > s \mid  S_i \geq s, B_i(s) \} = \prod_{s < t} \hat P\{e_{i}(s)=0 \mid  e_{i}(s^{-})=0, B_{i}(s) \}$, where $(e_{i}(s), B_{i}(s)) \in \mathcal{V}$.
\end{enumerate}

The probability $P\{S_i \geq t \mid  B_i(t)\}$ can also be estimated using a Cox proportional hazards model, where the event is defined by $e_{i}(t)=1$. In this context, $P\{S_i \geq t \mid  B_i(t)\}$ represents the probability that the subject $i$ ``survived'' until time $t$ given $B(t)$, and can be readily obtained using standard software. \cite{graffeo2019ipcwswitch} illustrate how to estimate such weights via a Cox model and provide software for implementation.

Existing literature suggests that when the time intervals are sufficiently small, and $P\{e_{i}(t)=1 \mid  e_{i}(t^{-})=0\}$ is low within each interval, the pooled logistic regression yields results virtually identical to the Cox model \citep{d1990relation}. However, it is important to acknowledge that this statement implicitly relies on the assumption that $L(t)$ is measured as frequently as it updates. In practice, measurements of $L(t)$ occur periodically, and any missing values between measurements are typically imputed using the last observed value. The estimated $\hat P\{S_i \geq t \mid  B_i(t)\}$ from the Cox model is a stepwise function. This stepwise nature does not accurately approximate the assumed smooth and continuous underlying switching process. In contrast, the pooled logistic regression derives $\hat P\{S_i \geq t \mid  B_i(t)\}$ through a cumulative product: $\hat P\{S_i \geq t \mid  B_i(t)\} = \prod_{s < t} \hat P\{e_{i}(s)=0 \mid  e_{i}(s^{-})=0, B_{i}(s) \}$. This approach ensures that the estimated weights are updated as frequently as $L(t)$ is updated, providing a smoother approximation of the underlying switching process. 

Suppose the estimation model for $P\{S_i \geq t \mid  B_i(t)\}$ is correct, $\hat P\{S_i \geq t \mid  B_i(t)\}$ converges to $P\{S_i \geq t \mid  B_i(t)\}$ with larger sample sizes. We can thus construct an asymptotically valid estimator by replacing $P\{S_i \geq t \mid  B_i(t)\}$ with $\hat P\{S_i \geq t \mid  B_i(t)\}$. 

For numerical stability and statistical efficiency, it is common practice to use stabilized weights \citep{cole2008constructing}, such as $\frac{\hat P(S_i \geq t \mid  X)}{\hat P\{S_i \geq t \mid  B(t)\}}$, as opposed to unstabilized weights $\frac{1}{\hat P\{S_i \geq t \mid  B(t)\}}$. Stabilized weights are employed to mitigate extreme weights. When utilizing stabilized weights, it becomes necessary to directly incorporate $X$ into the recurrent event model \citep{cole2008constructing}. This can be elucidated as follows: If $X$ is correlated with the switching process, while $L(t)$ is not, the stabilized weights $\frac{\hat P(S_i \geq t \mid  X)}{\hat P\{S_i \geq t \mid  B(t)\}}$ simplify to $\frac{\hat P(S_i \geq t \mid  X)}{\hat P(S_i \geq t \mid  X)}=1$. Thus, the baseline $X$ is not accounted for by weights, and we need to include $X$ directly in the recurrent event model. As a result, the estimated effect will be conditional on the baseline covariates rather than unconditional (marginal) \citep{cole2008constructing}.

One caveat concerning the variance of IPW-based estimators is that standard software treats the weights as fixed, disregarding their inherent randomness. In most cases, valid variance estimates can be obtained through bootstrap methods \citep{austin2016variance, tibshirani1993introduction}. To account for the uncertainty introduced by estimating $w_i(t)$, we construct bootstrap confidence intervals (CIs) as follows:
\begin{enumerate}
    \item Resample subjects with replacement to create a bootstrap sample of the same size as the original dataset. Since resampling may select the same subject multiple times, we assign a unique new subject ID to each sampled record, treating each as an independent subject for bootstrap purposes.
    \item Estimate $\beta$ from the bootstrap sample by first re-estimating $w_i(t)$ within the bootstrap sample and then fitting the LWYY + IPW, NB + IPW, and Na{\"i}ve NB + IPW models using the resulting $\hat{w}_i(t)$.
    \item Repeat steps 1 and 2 for a total of $B$ bootstrap replicates. Let $\{\hat \beta_{(1)}, \hat \beta_{(2)}, \dots, \hat \beta_{(B)}\}$ denote the resulting bootstrap estimates. Define $\beta^{*}_{(\alpha/2)}$ and $\beta^{*}_{(1-\alpha/2)}$ as the empirical $100(\alpha/2)\%$ and $100(1-\alpha/2)\%$ quantiles of these estimates. The $100(1 - \alpha)\%$ bootstrap CI is then $(\beta^{*}_{(\alpha/2)}, \beta^{*}_{(1-\alpha/2)})$, using the percentile method \citep{efron1982jackknife}.
\end{enumerate}

\section{Simulation studies}\label{sec:simulation}

\subsection{Overview}
The simulation considers a trial that involves a total of 2000 subjects, with an allocation ratio of 1:1. The trial duration spans 4 years, including 2 years of uniform enrollment. In the absence of competing risks of death, subjects in the trial experienced either independent loss to follow-up (i.e. when subjects fail to return for follow-up evaluations) or administrative censoring (i.e. when subjects reach a prespecified follow-up time). The follow-up time is modeled to follow an exponential distribution, which is subject to truncation by administrative censoring. Around 9\% of subjects are lost to follow-up. 

In modeling the time-varying covariate $L_i(t)$, we consider a single continuous variable. At baseline $t=0$, we initialize $L_i(0)$ by drawing from a normal distribution with parameters following a mean of 18 and a standard deviation of 5. For the baseline covariates $X_i$, we consider three variables in addition to the treatment assignment: sex, age, and prior disease history. To simulate sex, we employ a Bernoulli distribution with a probability of 0.5. Age is generated from a uniform distribution spanning the range of 50 to 65. The prior disease history is also generated from a Bernoulli distribution: if a subject's $L_{i}(0)$ value exceeds 16, the probability of prior disease history is set at 0.1; otherwise, it remains at 0.05. 

\subsection{Time-varying covariate}
The time unit for generating $L_i(t)$ is one week. We assume all subjects can be categorized into L-responders and L-nonresponders that have different time profiles of $L(t)$. For subject $i$, if the baseline value $L_{i}(0) < 15$, they are classified as L-nonresponders, while if $L_{i}(0)\geq 15$, they have an 80\% chance of being an L-responder and a 20\% chance of being an L-nonresponder. For L-nonresponders, we assume $L_{i}(t) \sim N(L_{i}(0),1)$ for $t>0$, meaning that $L_{i}(t)$ only has random fluctuations around the baseline value. For L-responders in the treatment arm, at time $t$, if $L_{i}(0)-0.14t<15$, then $L_{i}(t) \sim N(15,1)$; if $L_{i}(0)-0.14t \geq 15$, then $L_{i}(t) \sim N(L_{i}(0)-0.14t,1)$. In other words, for L-responders in the treatment arm, $L_i(t)$ decreases over time and stabilizes around 15. 

The correlation between $L_i(t)$ and $L_i(t')$ for subject $i$ ($t \neq t'$) is induced by their shared dependence on $L_i(0)$. Although we do not explicitly use a shared frailty model, our simulation follows a similar structure: post-randomization outcomes are functions of a shared baseline that is independent of treatment and fixed over time. Here, $L_i(0)$ plays the role of the shared frailty, with $L_i(t)$ and $L_i(t')$ defined as functions of $L_i(0)$, thereby inducing correlation.

In practical settings, when a subject decides to switch treatment, they may transition to the test treatment or to an alternative treatment available in the market. In our simulations, we follow a simplified protocol: switchers in the placebo arm switch to the treatment arm, while switchers in the treatment arm remain in the same arm. Therefore, for switchers in the treatment arm, their switching decision has no impact on the trajectory of $L(t)$. For subjects in the placebo arm, $L_{i}(t)$ follows $N(L_{i}(0),1)$ for $t>0$. This indicates that the values of $L(t)$ tend to remain stable around their respective baseline values. However, when an L-responder in the placebo arm switches treatment at time $t_{i}^{s}$, their $L_{i}(t)$ trajectory starts aligning with the treatment arm from $t_{i}^{s}$. Specifically, for $t \geq t_{i}^{s}$, if $L_{i}(0)-0.14  (t-t_{i}^{s})<15$, then $L_{i}(t) \sim N(15,1)$; if $L_{i}(0)-0.14  (t-t_{i}^{s}) \geq 15$, then $L_{i}(t) \sim N(L_{i}(0)-0.14  (t-t_{i}^{s}),1)$.

\begin{figure}[htbp]
    \centering
    \includegraphics[width=0.95\textwidth]{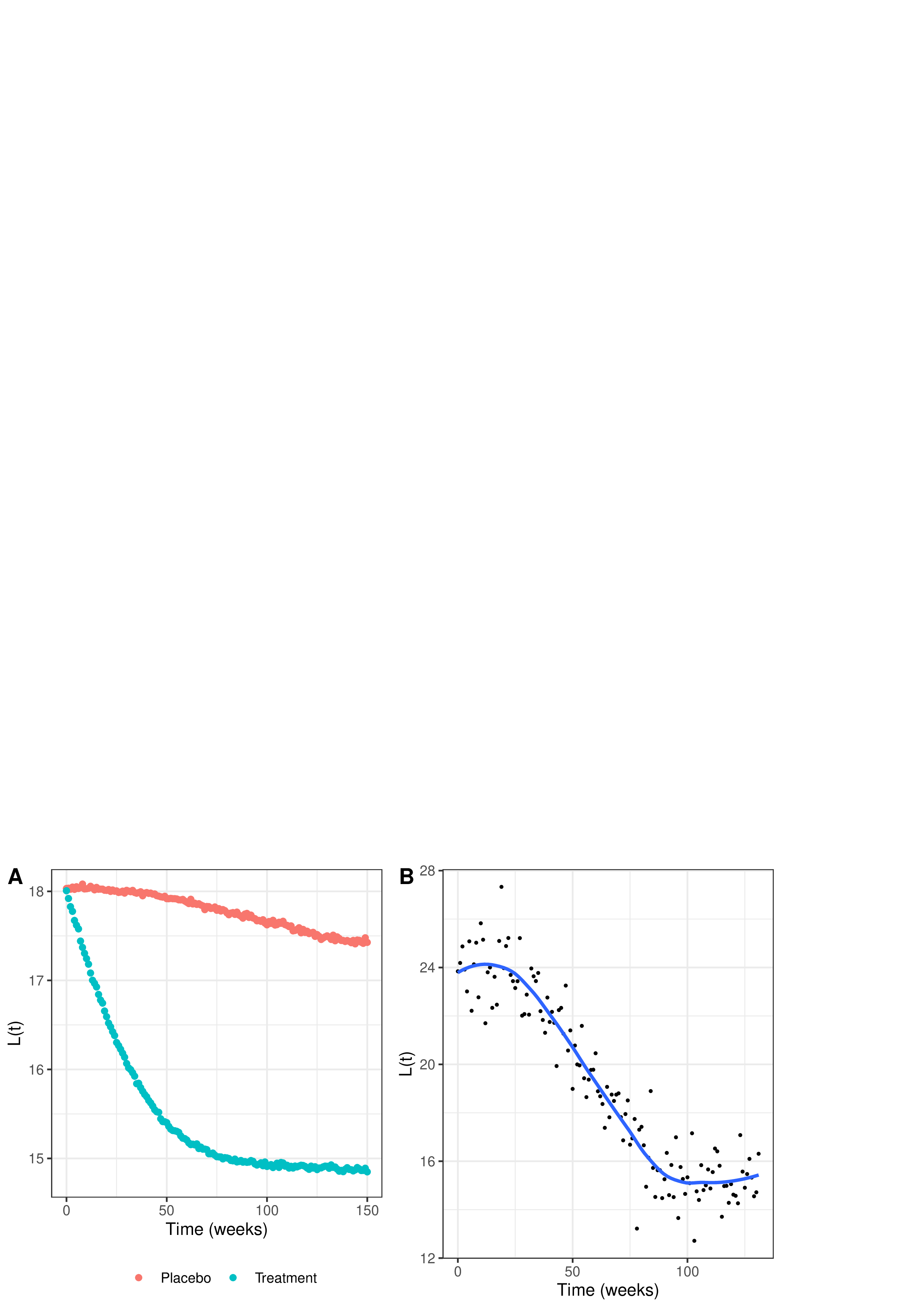}
    \caption{The trajectory of the time-varying covariate at the population level (left) and subject level (right).}
    \label{fig:lt}
\end{figure}

In Figure \ref{fig:lt}, Panel A illustrates the population-level trajectory of $L(t)$. At each time point, the average of $L(t)$ values from subjects at risk is calculated. A total of 10000 subjects are included to produce a smoother curve. In the treatment arm, $L(t)$ initially decreases but eventually stabilizes around 15 as time progresses whereas in the placebo arm, $L(t)$ initially remains relatively constant, but with a slight declining trend due to some subjects switching treatments. Panel B of Figure \ref{fig:lt} illustrates the trajectory of $L(t)$ for a particular subject, with the blue curve representing a smoothed line generated using the loess method. This subject, an L-responder, initially had a relatively high value of $L(0)$ and was assigned to the placebo arm. At week 28, the subject switched the treatment. Then $L(t)$ began to decrease under the treatment but subsequently stabilized around 15.

While in our simulations, $L(t)$ is generated and updated on a weekly basis, in practical settings, $L(t)$ is typically measured periodically. We explore two measurement scenarios. In the first scenario, we measure $L(t)$ weekly, matching the frequency of its updates. In the second scenario, $L(t)$ is assessed every 12 weeks. In cases where measurements are not available, values of $L(t)$ between two measurement time points are imputed using the ``last observation carried forward'' approach.

\subsection{Switching mechanism}
We implement a weekly switching mechanism, wherein the decision of whether a subject switches treatment on a given week is modeled as a Bernoulli event with the probability denoted as $p_{i}(t)=\left[1+\hbox{exp}\left\{-B_{i}^{T}(t) \beta_{s} \right\} \right]^{-1}$. The vector $B_{i}(t)$ includes treatment assignment, sex, age, prior disease history, and the time-varying covariate $L_i(t)$. The choice of $\beta_{s}$ is shown in Table \ref{simpara}. We assume no more than one treatment switching occurs for each subject in our simulations. The time-varying covariate $L_{i}(t)$ and the switching probability $p_{i}(t)$ exhibit a positive correlation in our simulation setup.

\subsection{Recurrent event mechanism}
We also generate recurrent events on a weekly basis, considering three scenarios. In the first scenario, we model whether a subject experiences an event on a given week as a Bernoulli event, with a probability denoted as $q_{i}^{(1)}(t) = \left[1+ \exp\left\{-B_{i}^{T}(t) \beta_{e}^{(1)} \right\}\right]^{-1}$. In this scenario, $q_{i}^{(1)}(t)$ does not depend on the event history. In the second scenario, we define the indicator $E_{i}(t)$, which tracks whether subject $i$ has ever experienced recurrent events up to time $t$. Events are still generated on a weekly basis, but the probability now depends on $E_{i}(t)$, resulting in $q_{i}^{(2)}(t) = \left[1+ \exp\left\{-B_{i}^{T}(t) \beta_{e}^{(2)} + 0.7  E_{i}(t)\right\}\right]^{-1}$. In the third scenario, we assign a random variable $\gamma_i$ to each subject following a gamma distribution with the mean as 1 and the variance as 0.5. Events are again generated on a weekly basis with the Bernoulli probability $q_{i}^{(3)}(t) = \gamma_i  \left[1+ \exp\left\{-B_{i}^{T}(t) \beta_{e}^{(3)}\right\}\right]^{-1}$. If $q_{i}^{(3)}(t) >1$, we bound it by 1. Henceforth, we refer to these scenarios as ``Scenario 1,'' ``Scenario 2,'' and ``Scenario 3,'' respectively. Note that $L_{i}(t)$ is positively correlated with the event probability across all three scenarios. The choices of $\beta_{e}^{(1)}$, $\beta_{e}^{(2)}$, and $\beta_{e}^{(3)}$ are shown in Table \ref{simpara}.

When fitting LWYY and NB models, the covariates include treatment assignment, sex, age, and prior disease history. With the IPW approach to account for $L(t)$, the model forms for LWYY + IPW and NB + IPW do not incorporate the previous event history indicator $E_{i}(t)$ in Scenario 2, nor do they account for the per-subject random variable $\gamma_i$ in Scenario 3. Therefore, Scenarios 2 and 3 serve as robustness checks of our approaches.

Since our focus is on the estimand in a hypothetical scenario where no switching occurs, to generate the empirical true value and then calculate the bias of different analysis approaches, an additional variable $L^{*}(t)$ is generated in each simulated dataset. $L^{*}(t)$ represents $L(t)$ in the hypothetical scenario where no switching occurs, differing from $L(t)$ only after the switching time $t^{s}$ among L-responders initially assigned to the placebo arm. The hypothetical recurrent event process in the absence of treatment switching is then simulated using the same formulas $q_{i}^{(1)}(t)$, $q_{i}^{(2)}(t)$, and $q_{i}^{(3)}(t)$ but substituting $L(t)$ with $L^{*}(t)$.

\subsection{Results}
We compare different estimation methods for the hypothetical estimand of interest, including a simple censoring approach (treating treatment switching as independent censoring and analyzing only events occurring before switching for each subject) and the proposed IPW approaches. We also present the results from analyses targeting the treatment policy estimand, which analyzes all events based on their assigned arm, regardless of treatment switching. For each approach considered, both the LWYY and NB models are fitted. The weights are estimated using the pooled logistic regression model, as described in Section~\ref{weightest}.

Table \ref{datasummary} presents a summary of key statistics for the generated datasets obtained from 1000 simulations under three scenarios. The average event rate in the placebo arm under the treatment policy estimand is lower than that in the hypothetical scenario due to lower event risk among switchers in the placebo arm. With the simple censoring approach, the average follow-up time of the placebo arm is shorter compared to that of the treatment arm. This disparity arises because $L(t)$ is positively correlated with the probability of switching, and on average, $L(t)$ is higher in the placebo arm.

Tables \ref{resultssce1}, \ref{resultssce2}, and \ref{resultssce3} present estimation results from various models and approaches for Scenarios 1 to 3, respectively, each based on 1000 simulations. ``Est'' represents the Monte Carlo sample mean of $\hat{\beta}$. The ``Est'' from the hypothetical scenario, where no switching occurred, serves as the empirical true value of the corresponding model. ``SD'' denotes the standard deviation of $\hat{\beta}$. ``Bias'' indicates the Monte Carlo sample mean of $\hat{\beta}$ minus the corresponding empirical true value. The rate ratio (RR) is computed as the exponential of the corresponding ``Est''. Coverage probability (CP) refers to the likelihood that the 95\% confidence interval (CI) contains the empirical true value of the corresponding model. The CIs for the hypothetical scenario, simple censoring approach, and treatment policy approach are based on the asymptotic normal approximation, while the CIs for IPW approaches are derived from the percentile (2.5th and 97.5th percentiles) interval of 1000 bootstrap estimates \citep{efron1982jackknife}.

Note that the treatment not only has direct impacts on the recurrent event process but also has indirect effects through $L(t)$, suggesting that the true rate ratio is time-dependent rather than strictly constant. The constant rate ratio estimated from LWYY and NB models is essentially a weighted sum of treatment effects over time, which, however, still serves as an informative summary of treatment effects.

The results remain consistent across the three scenarios. Whether measuring $L(t)$ weekly or at 12-week intervals, the estimation results show minimal variation. The LWYY and NB models yield nearly identical rate ratio estimates in the hypothetical scenario with no treatment switching. Similarly, in the presence of treatment switching, when applying the same analytic approach---such as treatment policy, simple censoring, or IPW---the two models yield similar rate ratio estimates. The biases of LWYY + IPW, NB + IPW, and Na{\"i}ve NB + IPW are minimal, likely due to random variability. Estimates of the rate ratio from the simple censoring and treatment policy approaches are closer to 1 compared to the IPW-based approaches. The IPW approach shows coverage probabilities that closely align with the nominal level, indicating well-controlled Type I error.

Table \ref{resultspower} presents the power of LWYY and NB models using various approaches in multiple scenarios, each involving 1000 simulations. Power is calculated using a significance level of $\alpha=0.05$ for a two-sided test. Strictly speaking, when referring to power, we are comparing the probability of rejecting the null hypothesis---that the treatment has no effect---across different approaches, while acknowledging that the null hypotheses for each approach may differ. The ``robust'' (R) method treats weights as fixed, while the ``bootstrap'' (B) method considers the variability of weights, each utilizing 1000 bootstraps. Estimates from the hypothetical scenario serve as a reference for comparison. Power remains similar when measuring $L(t)$ weekly or at 12-week intervals, and consistent across all scenarios. For the IPW approach, the bootstrap method shows comparable power to the robust method. The power of LWYY + IPW and NB + IPW are comparable, with both showing slightly higher power than Na{\"i}ve NB + IPW. The power of the simple censoring and treatment policy approaches are comparable to each other, but lower than that of all IPW approaches.

\section{Re-analysis of synthetic trial data}
\label{sec:asthmaanalysis}
In this section, we illustrate the different statistical methods for estimating hypothetical estimands using synthetic data from the Phase III clinical trial described in Section \ref{sec:example:asthma}.
The synthetic dataset includes four baseline covariates: region, history of asthma exacerbations, forced expiratory volume in 1 second (FEV1) before inhalation, and FEV1 after inhalation. Data collection for some patients was completed before the planned Week 26 due to administrative reasons, referred to as ``administrative censoring.'' Additionally, no data is available for the period after patients discontinued their randomized treatment, meaning all observed data is on-treatment data. Patients could either discontinue their randomized treatment or be censored for administrative reasons. Table \ref{tab:asthmaoverview} provides the number of patients randomized, those with the intercurrent event for each trial arm as well as the mean number of events and the mean number of days on the randomized treatment.

We applied the IPW approaches (LWYY + IPW, NB + IPW, Na{\"i}ve NB + IPW) to this synthetic dataset and also included results from LWYY and NB with simple censoring for comparison. The weights are estimated using the pooled logistic regression model described in Section~\ref{weightest}. The four baseline covariates described above were included in all models. For the IPW approaches, we incorporated two time-varying covariates: evening peak expiratory flow (PEF) and the total daily number of puffs of rescue medication, both measured daily. The analysis results are presented in Table \ref{tab:asthmaresults}. Bootstrap standard errors (SEs) are not provided for models with simple censoring, as no estimated weights are involved, making the computation of bootstrap SEs unnecessary. Robust SEs are unavailable for NB + IPW because this model was implemented using our own code rather than standard software. The results from the simple censoring approach and IPW are similar, possibly due to the low and balanced intercurrent event rate in both arms (about 7\% per arm). In addition, the two time-varying covariates may not fully capture all different reasons leading to treatment discontinuation, so the added prediction value from the time-varying covariates on top of other existing covariates may be limited. Note that the analysis of the synthetic data is for illustration purposes only, so the results should be interpreted with caution.

\section{Conclusion}\label{sec:conclusion}

In this paper, we presented statistical methods for the estimation of hypothetical estimands in randomized trials such as ``What would the treatment effect be if subjects had not experienced the intercurrent event?'' Estimating these estimands is complicated by the presence of time-varying confounders that influences both intercurrent and recurrent event processes. Proper adjustment for time-varying confounders is crucial; however, directly including the time-varying confounder as a regressor is problematic. To resolve this dilemma, we employ the IPW approach, which involves artificially censoring subjects at the time of the intercurrent event, disregarding events that occur afterward, and assigning time-dependent weights based on the inverse probability of remaining free from the intercurrent event.

LWYY and NB models can be derived from a unified framework. Within this framework, we apply the IPW to each model and propose three approaches: LWYY + IPW, NB + IPW, and Na{\"i}ve NB + IPW. The weights can be estimated through a pooled logistic regression model. The three approaches differ in how they handle treatment switching in the following ways: First, while both NB + IPW and LWYY + IPW use all subjects in the estimating equation, Na{\"i}ve NB + IPW only includes non-switchers. Second, for each subject $i$ included in the estimating equations, Na{\"i}ve NB + IPW only use one weight $w_{i}(\tau_i)$; LWYY + IPW uses $w_{i}(t)$ specifically at recurrent event time points; and NB + IPW incorporates all $w_{i}(t)$ over $(0,\min(\tau_i, S_i))$. As a result, NB + IPW and LWYY + IPW tend to yield slightly higher power than Na{\"i}ve NB + IPW, making them more favorable choices over Na{\"i}ve NB + IPW.

Our simulation results should be interpreted with caution. First, as suggested by \cite{metcalfe2006importance}, simulation studies of statistical methods for recurrent events should include datasets based on a variety of event generation models. We only explore three possible recurrent event mechanisms. Second, we make the assumption that the intercurrent event on each week depends on observed covariates, which corresponds to a Missing at Random (MAR) assumption \citep{little2019statistical}.  Under MAR, when the intercurrent event mechanism is correctly specified, weighted estimating equations provide consistent estimation. However, when the intercurrent event mechanism is misspecified, weighted estimating equations might perform worse than the unweighted version \citep{preisser2002performance}. In our simulations, weights are derived from models aligned with the true intercurrent event mechanism. Nevertheless, in reality, the true intercurrent event mechanism is unknown and cannot be observed from the data.


\bibliography{maintext}  

\clearpage

\clearpage

\begin{table}[htbp]
\caption{The choice of coefficients $\beta_{s}$, $\beta_{e}^{(1)}$, $\beta_{e}^{(2)}$, and $\beta_{e}^{(3)}$ in the simulation.\label{simpara}}
\centering
\tabcolsep=8pt
\begin{tabular}{lcccccc}
\toprule
 & Intercept & \begin{tabular}[c]{@{}l@{}}Treatment Assignment\end{tabular} & \begin{tabular}[c]{@{}l@{}}Prior Disease History\end{tabular} & Sex & Age & $L(t)$ \\ \midrule
$\beta_{s}$       & -13.76 & -0.4  & 0.8   & 0.4  & 0.016  & 0.264 \\
$\beta_{e}^{(1)}$ & -5.6  & -0.07 & 0.07 & 0.035 & 0.0035 & 0.028 \\
$\beta_{e}^{(2)}$ & -5.74    & -0.07 & 0.07 & 0.035 & 0.0035 & 0.028 \\
$\beta_{e}^{(3)}$ & -5.46    & -0.105 & 0.07 & 0.035 & 0.0035 & 0.028 \\
\bottomrule
\end{tabular}
\begin{tablenotes}
\item[] \textbf{Notes:}
\begin{itemize}
    \item $\beta_{s}$ is the coefficient for generating the switching process.
    \item $\beta_{e}^{(1)}$, $\beta_{e}^{(2)}$, and $\beta_{e}^{(3)}$ correspond to coefficients for generating the recurrent event processes in the first, second, and third scenarios, respectively. 
\end{itemize}
\end{tablenotes}
\end{table}

\begin{table}[htbp]
\caption{Summary of key statistics for the generated datasets obtained from 1000 simulations under three scenarios.\label{datasummary}}
\centering
\tabcolsep=1pt
\begin{tabular}{ccccccccccccccc}
\toprule
 & & \multicolumn{4}{c}{Hypothetical} & \multicolumn{4}{c}{Treatment policy} & \multicolumn{5}{c}{Simple censoring} \\ \cmidrule(lr){3-6} \cmidrule(lr){7-10} \cmidrule(lr){11-15} 
Arm &
  \begin{tabular}[c]{@{}c@{}}\# \\ subjects\end{tabular} &
  \begin{tabular}[c]{@{}c@{}}\# \\ events\end{tabular} &
  \begin{tabular}[c]{@{}c@{}}Time \\ (year)\end{tabular} &
  \begin{tabular}[c]{@{}c@{}}Event rate \\ (per year)\end{tabular} &
  \begin{tabular}[c]{@{}c@{}}Average \\ \# events\end{tabular} &
  \begin{tabular}[c]{@{}c@{}}\# \\ events\end{tabular} &
  \begin{tabular}[c]{@{}c@{}}Time \\ (year)\end{tabular} &
  \begin{tabular}[c]{@{}c@{}}Event rate \\ (per year)\end{tabular} &
  \begin{tabular}[c]{@{}c@{}}Average \\ \# events\end{tabular} &
  \begin{tabular}[c]{@{}c@{}}\% \\ switcher\end{tabular} &
  \begin{tabular}[c]{@{}c@{}}\# \\ events\end{tabular} &
  \begin{tabular}[c]{@{}c@{}}Time \\ (year)\end{tabular} &
  \begin{tabular}[c]{@{}c@{}}Event rate \\ (per year)\end{tabular} &
  \begin{tabular}[c]{@{}c@{}}Average \\ \# events\end{tabular} \\ \midrule
\multicolumn{15}{l}{\textbf{Scenario 1}}                                                                                       \\
0 & 1000 & 1152 & 2.87 & 0.401 & 1.15 & 1136 & 2.87 & 0.395 & 1.14 & 11.7 & 1061 & 2.68 & 0.396 & 1.06 \\
1 & 1000 & 996 & 2.87 & 0.347 & 1.00 & 996 & 2.87 & 0.347 & 1.00 & 4.0 & 968 & 2.80 & 0.346 & 0.97 \\
\multicolumn{15}{l}{\textbf{Scenario 2}}                                                                                   \\
0 & 1000 & 1388 & 2.87 & 0.483 & 1.39 & 1364 & 2.87 & 0.475 & 1.36 & 11.7 & 1265 & 2.68 & 0.472 & 1.27 \\
1 & 1000 & 1167 & 2.87 & 0.406 & 1.17 & 1167 & 2.87 & 0.406 & 1.17 & 4.0 & 1131 & 2.80 & 0.404 & 1.13 \\ 
\multicolumn{15}{l}{\textbf{Scenario 3}}                                                                                   \\
0 & 1000 & 1323 & 2.87 & 0.461 & 1.32 & 1303 & 2.87 & 0.454 & 1.30 & 11.7 & 1219 & 2.68 & 0.455 & 1.22 \\
1 & 1000 & 1106 & 2.87 & 0.385 & 1.11 & 1106 & 2.87 & 0.385 & 1.11 & 4.0 & 1075 & 2.80 & 0.384 & 1.08 \\ \bottomrule
\end{tabular}
\begin{tablenotes}
\footnotesize
\item[] \textbf{Notes:}
\begin{itemize}
    \item The ``hypothetical'' scenario represents a hypothetical case in which no treatment switching occurred.   
    \item Arm (0 and 1) refers to the two study arms, with ``Arm 0'' representing the placebo arm and ``Arm 1'' representing the treatment arm.
    \item \# events indicates the total number of recurrent events experienced by all subjects.
    \item Time (Year) represents the average follow-up time of each subject in years.
    \item Event rate denotes the average number of events experienced by each subject per year.
    \item Average \# events refers to the average number of events experienced by each subject throughout their entire follow-up time.
    \item \% Switcher represents the average proportion of subjects who switched treatments.
\end{itemize}
\end{tablenotes}
\end{table}

\begin{table}[htbp]
\caption{Estimation results for LWYY and NB models utilizing various approaches in Scenario 1, based on 1000 simulations.\label{resultssce1}}
\centering
\tabcolsep=7pt
\begin{tabular}{lcccccccccc} \toprule
  & \multicolumn{5}{c}{$L^{1}(t)$} &
  \multicolumn{5}{c}{$L^{12}(t)$} \\ 
  \cmidrule(lr){2-6} \cmidrule(lr){7-11} 
&
  \multicolumn{1}{c}{Est} &
  \multicolumn{1}{c}{SD} &
  \multicolumn{1}{c}{Bias} &
  \multicolumn{1}{c}{RR} &
  \multicolumn{1}{c}{CP} &
  \multicolumn{1}{c}{Est} &
  \multicolumn{1}{c}{SD} &
  \multicolumn{1}{c}{Bias} &
  \multicolumn{1}{c}{RR} &
  \multicolumn{1}{c}{CP} \\ 
  \midrule
\multicolumn{11}{l}{\textbf{LWYY}} \\
Hypothetical &
-0.146 & 0.044 & 0.000 & 0.864 & 94.5 &
-0.146 & 0.044 & 0.000 & 0.864 & 94.5
\\ \\
LWYY + IPW &
-0.144 & 0.046 & 0.002 & 0.866  & 94.3 &
-0.144 & 0.046 & 0.002 & 0.866 & 94.5
\\ \\
Simple censoring &
-0.136 & 0.045 & 0.010 & 0.873 & 93.6 &
-0.136 & 0.045 & 0.010 & 0.873 & 93.6 
\\ \\
Treatment policy &
-0.131 & 0.044 & 0.014 & 0.877 & 93.0 &
-0.131 & 0.044 & 0.014 & 0.877 & 93.0
  \\  \midrule
\multicolumn{11}{l}{\textbf{NB}} \\
Hypothetical &
-0.145 & 0.044 & 0.000 & 0.865 & 94.7 &
-0.145 & 0.044 & 0.000 & 0.865 & 94.7
  \\ \\
NB + IPW &
-0.144 & 0.046 & 0.001 & 0.866 & 94.4 &
-0.144 & 0.046 & 0.001 & 0.866 & 94.5
  \\ \\
Na{\"i}ve NB + IPW &
-0.143 & 0.048 & 0.002 & 0.866 & 93.9 &
-0.143 & 0.048 & 0.002 & 0.867 & 94.1
  \\ \\
Simple censoring &
-0.136 & 0.045 & 0.009 & 0.873 & 93.9 &
-0.136 & 0.045 & 0.009 & 0.873 & 93.9
  \\ \\
Treatment policy &
-0.131 & 0.044 & 0.014 & 0.877 & 93.3 &
-0.131 & 0.044 & 0.014 & 0.877 & 93.3
  \\
\bottomrule 
\end{tabular}
\begin{tablenotes}
\footnotesize
\item[] \textbf{Notes:}
\begin{itemize}
    \item $L^{1}(t)$ denotes $L(t)$ being updated every week.
    \item $L^{12}(t)$ denotes $L(t)$ being updated every 12 weeks.
\end{itemize}
\end{tablenotes}
\end{table}

\begin{table}[htbp]
\caption{Estimation results for LWYY and NB models utilizing various approaches in Scenario 2, based on 1000 simulations.    \label{resultssce2}}
\centering
\tabcolsep=7pt
\begin{tabular}{lccccccccccc} \toprule
  & \multicolumn{5}{c}{$L^{1}(t)$} &
  \multicolumn{5}{c}{$L^{12}(t)$} \\ 
  \cmidrule(lr){2-6} \cmidrule(lr){7-11} 
&
  \multicolumn{1}{c}{Est} &
  \multicolumn{1}{c}{SD} &
  \multicolumn{1}{c}{Bias} &
  \multicolumn{1}{c}{RR} &
  \multicolumn{1}{c}{CP} &
  \multicolumn{1}{c}{Est} &
  \multicolumn{1}{c}{SD} &
  \multicolumn{1}{c}{Bias} &
  \multicolumn{1}{c}{RR} &
  \multicolumn{1}{c}{CP} \\ 
  \midrule
\multicolumn{11}{l}{\textbf{LWYY}} \\
Hypothetical &
-0.174 & 0.050 & 0.000 & 0.840 & 93.8 &
-0.174 & 0.050 & 0.000 & 0.840 & 93.8
\\ \\
LWYY + IPW &
-0.172 & 0.053 & 0.002 & 0.842  & 94.7 &
-0.172 & 0.052 & 0.002 & 0.842 & 94.5
\\ \\
Simple censoring &
-0.161 & 0.051 & 0.013 & 0.851 & 93.0 &
-0.161 & 0.051 & 0.013 & 0.851 & 93.0 
\\ \\
Treatment policy &
-0.156 & 0.050 & 0.018 & 0.855 & 93.0 &
-0.156 & 0.050 & 0.018 & 0.855 & 93.0
  \\  \midrule
\multicolumn{11}{l}{\textbf{NB}} \\
Hypothetical &
-0.172 & 0.050 & 0.000 & 0.842 & 94.8 &
-0.172 & 0.050 & 0.000 & 0.842 & 94.8
  \\ \\
NB + IPW &
-0.171 & 0.052 & 0.001 & 0.843 & 94.1 &
-0.170 & 0.052 & 0.002 & 0.843 & 94.1
  \\ \\
Na{\"i}ve NB + IPW &
-0.169 & 0.054 & 0.003 & 0.845 & 93.8 &
-0.168 & 0.054 & 0.004 & 0.845 & 93.7
  \\ \\
Simple censoring &
-0.156 & 0.051 & 0.016 & 0.855 & 93.6 &
-0.156 & 0.051 & 0.016 & 0.855 & 93.6
  \\ \\
Treatment policy &
0.155 & 0.050 & 0.017 & 0.856 & 94.0 &
-0.155 & 0.050 & 0.017 & 0.856 & 94.0
  \\
\bottomrule 
\end{tabular}
\begin{tablenotes}
\footnotesize
\item[] \textbf{Notes:}
\begin{itemize}
    \item $L^{1}(t)$ denotes $L(t)$ being updated every week.
    \item $L^{12}(t)$ denotes $L(t)$ being updated every 12 weeks.
\end{itemize}
\end{tablenotes}
\end{table}

\begin{table}[htbp]
\caption{Estimation results for LWYY and NB models utilizing various approaches in Scenario 3, based on 1000 simulations. \label{resultssce3}}
\centering
\tabcolsep=7pt
\begin{tabular}{lccccccccccc} \toprule
  & \multicolumn{5}{c}{$L^{1}(t)$} &
  \multicolumn{5}{c}{$L^{12}(t)$} \\ 
  \cmidrule(lr){2-6} \cmidrule(lr){7-11} 
&
  \multicolumn{1}{c}{Est} &
  \multicolumn{1}{c}{SD} &
  \multicolumn{1}{c}{Bias} &
  \multicolumn{1}{c}{RR} &
  \multicolumn{1}{c}{CP} &
  \multicolumn{1}{c}{Est} &
  \multicolumn{1}{c}{SD} &
  \multicolumn{1}{c}{Bias} &
  \multicolumn{1}{c}{RR} &
  \multicolumn{1}{c}{CP} \\ 
  \midrule
\multicolumn{11}{l}{\textbf{LWYY}} \\
Hypothetical &
-0.180 & 0.053 & 0.000 & 0.835 & 94.7 &
-0.180 & 0.053 & 0.000 & 0.835 & 94.7
\\ \\
LWYY + IPW &
-0.180 & 0.055 & 0.001 & 0.835  & 94.8 &
-0.179 & 0.055 & 0.001 & 0.836 & 94.7
\\ \\
Simple censoring &
-0.171 & 0.054 & 0.009 & 0.843 & 94.5 &
-0.171 & 0.054 & 0.009 & 0.843 & 94.5 
\\ \\
Treatment policy &
-0.165 & 0.053 & 0.016 & 0.848 & 92.8 &
-0.165 & 0.053 & 0.016 & 0.848 & 92.8
  \\  \midrule
\multicolumn{11}{l}{\textbf{NB}} \\
Hypothetical &
-0.179 & 0.053 & 0.000 & 0.836 & 94.9 &
-0.179 & 0.053 & 0.000 & 0.836 & 94.9
  \\ \\
NB + IPW &
-0.178 & 0.054 & 0.001 & 0.837 & 94.9 &
-0.178 & 0.054 & 0.001 & 0.837 & 94.6
  \\ \\
Na{\"i}ve NB + IPW &
-0.179 & 0.057 & 0.001 & 0.836 & 94.4 &
-0.178 & 0.057 & 0.001 & 0.837 & 94.5
  \\ \\
Simple censoring &
-0.172 & 0.054 & 0.008 & 0.842 & 94.5 &
-0.172 & 0.054 & 0.008 & 0.842 & 94.5
  \\ \\
Treatment policy &
-0.164 & 0.052 & 0.015 & 0.849 & 93.8 &
-0.164 & 0.052 & 0.015 & 0.849 & 93.8
  \\
\bottomrule 
\end{tabular}
\begin{tablenotes}
\footnotesize
\item[] \textbf{Notes:}
\begin{itemize}
    \item $L^{1}(t)$ denotes $L(t)$ being updated every week.
    \item $L^{12}(t)$ denotes $L(t)$ being updated every 12 weeks.
\end{itemize}
\end{tablenotes}
\end{table}

\begin{table}[htbp]
\caption{Power analysis for LWYY and NB models utilizing various approaches in multiple scenarios, based on 1000 simulations. \label{resultspower}}
\centering
\tabcolsep=12pt
\begin{tabular}{lcccccc} \toprule
 &
  \multicolumn{2}{c}{Scenario 1} &
  \multicolumn{2}{c}{Scenario 2} & \multicolumn{2}{c}{Scenario 3} \\ 
  \cmidrule(lr){2-3} \cmidrule(lr){4-5} \cmidrule(lr){6-7}
 &
  \multicolumn{1}{c}{$L^{1}(t)$} &
  \multicolumn{1}{c}{$L^{12}(t)$} &
  \multicolumn{1}{c}{$L^{1}(t)$} &
  \multicolumn{1}{c}{$L^{12}(t)$} &
  \multicolumn{1}{c}{$L^{1}(t)$} & 
  \multicolumn{1}{c}{$L^{12}(t)$} \\ 
  \midrule
\multicolumn{7}{l}{\textbf{LWYY}} \\
Hypothetical &
  90.8 &
  90.8 &
  94.2 &
  94.2 &
  92.4 &
  92.4 \\ \\
LWYY + IPW &
  \begin{tabular}[c]{@{}l@{}}88.5 (R)\\ 88.8 (B) \end{tabular} &
  \begin{tabular}[c]{@{}l@{}}88.7 (R)\\ 88.7 (B) \end{tabular} &
  \begin{tabular}[c]{@{}l@{}}91.4 (R)\\ 91.4 (B) \end{tabular} &
  \begin{tabular}[c]{@{}l@{}}91.2 (R)\\ 91.3 (B) \end{tabular} &
  \begin{tabular}[c]{@{}l@{}}90.7 (R)\\ 90.3 (B) \end{tabular} &
  \begin{tabular}[c]{@{}l@{}}90.7 (R)\\ 90.4 (B) \end{tabular} \\ \\
Simple censoring &
  86.4 &
  86.4 &
  88.8 &
  88.8 &
  88.7 &
  88.7 \\ \\
Treatment policy &
  84.8 &
  84.8 &
  89.0 &
  89.0 &
  87.5 &
  87.5 \\   \midrule
\multicolumn{7}{l}{\textbf{NB}} \\
Hypothetical &
  90.9 &
  90.9 &
  93.1 &
  93.1 &
  92.5 &
  92.5 \\ \\
NB + IPW &
  88.6 (B) &
  88.6 (B) &
  91.3 (B) &
  91.2 (B) &
  90.3 (B) &
  90.6 (B) \\ \\
Na{\"i}ve NB + IPW &
  \begin{tabular}[c]{@{}l@{}} 88.7 (R) \\ 87.3 (B) \end{tabular} &
  \begin{tabular}[c]{@{}l@{}} 88.7 (R) \\ 87.4 (B) \end{tabular} &
  \begin{tabular}[c]{@{}l@{}} 89.6 (R) \\ 89.2 (B) \end{tabular} &
  \begin{tabular}[c]{@{}l@{}} 89.6 (R) \\ 89.5 (B) \end{tabular} &
  \begin{tabular}[c]{@{}l@{}} 89.7 (R) \\ 88.5 (B) \end{tabular} &
  \begin{tabular}[c]{@{}l@{}} 89.9 (R) \\ 88.7 (B) \end{tabular} \\ \\
Simple censoring &
  86.7 &
  86.7 &
  86.7 &
  86.7 &
  89.5 &
  89.5 \\ \\
Treatment policy &
  84.8 &
  84.8 &
  87.5 &
  87.5 &
  87.7 &
  87.7 \\
\bottomrule 
\end{tabular}
\begin{tablenotes}
\footnotesize
\item[] \textbf{Notes:}
\begin{itemize}
    \item $L^{1}(t)$ denotes $L(t)$ being updated every week.
    \item $L^{12}(t)$ denotes $L(t)$ being updated every 12 weeks.
\end{itemize}
\end{tablenotes}
\end{table}

\begin{table}[htbp]
\caption{Key characteristics of the synthetic asthma data.}
\label{tab:asthmaoverview}
\centering
\begin{tabular}{lcc}
\toprule
 & \textbf{Control} & \textbf{Drug X} \\
\midrule
Number of randomized patients & 606 & 612 \\

Number of patients with the intercurrent event & 39 & 41 \\

Mean number of exacerbations & 0.378 & 0.291 \\

Mean number of days on the randomized treatment & 169 & 171 \\
\bottomrule 
\end{tabular}

\end{table}

\clearpage

\begin{table}[htbp]
\caption{Analysis results of the synthetic asthma trial data.\label{tab:asthmaresults}}
\centering
\tabcolsep=12pt
\begin{tabular}{lccccc} \toprule
 & &  \multicolumn{2}{c}{$SE(\hat \beta)$} & \\ 
  \cmidrule(lr){3-4} 
 &
  \multicolumn{1}{c}{$\hat \beta$} &
  \multicolumn{1}{c}{Robust} &
  \multicolumn{1}{c}{Bootstrap} &
  \multicolumn{1}{c}{$\hbox{exp}(\hat \beta)$} &
  \multicolumn{1}{c}{$95\%$ CI} \\ 
  \midrule
\multicolumn{6}{l}{\textbf{LWYY}} \\
Simple censoring &
  -0.283 & 0.142 & $-$ & 0.753 & (0.570, 0.996) \\ \\
LWYY + IPW &
  -0.282 & 0.143 & 0.149 & 0.754 & (0.569, 1.003)\\ 
\midrule
\multicolumn{6}{l}{\textbf{NB}} \\
Simple censoring  &
  
  -0.268 & 0.137 & $-$ &0.765 & (0.585, 1.000) \\ \\
NB + IPW &
  
  -0.257 & $-$ & 0.144 &0.773 & (0.586, 1.015) \\ \\
  Na{\"i}ve NB + IPW &
  
  -0.221 & 0.147 & 0.155 &0.802 & (0.592, 1.086) \\
\bottomrule 
\end{tabular}
\begin{tablenotes}
\footnotesize
\item[] \textbf{Notes:}
\begin{itemize}
    \item The robust SE treats the estimated weights as fixed.
    \item  The bootstrap SE is the standard deviation of 1000 bootstrap estimates of $\beta$.
    \item $\hbox{exp}(\hat \beta)$ represents the estimated rate ratio.
    \item CIs are obtained from the 2.5th and 97.5th percentiles of the 1000 bootstrap estimates.
\end{itemize}
\end{tablenotes}
\end{table}

\end{document}


\maketitle
\section{Probability Density Function for an Event Process}\label{sec:appen:pdf}
In this section, we derive the probability density function for an event process that is observed over the fixed time interval $[\tau_0,\tau]$, conditional on $H(\tau_0)$ (Chapter 2 of \cite{cook2007statistical}). The probability density of the outcome ``$n$ events occur, at times $\tau_0 < t_1 < t_2 < \dots < t_n \leq \tau$,'' where $n \geq 0$, may be obtained by considering partitions $\tau_0 = u_0 < u_1 < \dots < u_R = \tau$ of $[\tau_0,\tau]$, and then taking a limit. The probability distribution of $\Delta N(u_1),\ldots, \Delta N(u_R)$ given $H(u_0)$ is 
\begin{eqnarray}
\prod_{r=0}^{R-1} P\{\Delta N(u_r) \mid  H(u_r) \} = \prod_{r=0}^{R-1}  \left[ \lambda\{u_r \mid  H(u_r)\}\Delta u_r + o(\Delta u_r) \right]^{\Delta N(u_r)} \nonumber \\ 
\times \left[ 1-\lambda\{u_r \mid  H(u_r)\}\Delta u_r + o(\Delta u_r)\right]^{1-\Delta N(u_r)}  \label{lld1}
\end{eqnarray}
where $\Delta N(u_r)$ is the number of events in $[u_r, u_{r+1})$, and $o(\Delta u_r)$ denotes a term tending to 0 as $\Delta u_r$ approaches zero.

As R increases and the size of the $\Delta u_r$ terms approach zero, the $n$ intervals that contain the event times $t_1, \dots, t_n$ have $\Delta N(u_r)=1$; for all others $\Delta N(u_r)=0$. By noting that $\hbox{log}\{ 1+g(t)\Delta t\}=g(t)\Delta t+o(\Delta t)$, where $g(t)$ is a continuous integrable function over $(\tau_0, \tau)$, Equation (\ref{lld1}) can be written as
\begin{eqnarray*}
  \prod_{r=0}^{R-1} P\{\Delta N(u_r) \mid  H(u_r) \} = \prod_{j=1}^{n} \lambda \{ t_j \mid  H(t_j) \}  \hbox{exp} \left[ -\int_{\tau_0}^{\tau} \lambda\{t \mid  H(t)\}dt  \right]  \label{lld2}.
\end{eqnarray*}

Let $Y(t)=I(\hbox{process is observed at time } t)$. If the at-risk process $Y(t)$ is conditionally independent of the event process $N(t)$, given the history $H(t)$, then we have
\begin{eqnarray*}
    \prod_{r=0}^{R-1} P\{\Delta N(u_r) \mid  H(u_r) \} = \prod_{j=1}^{n} \lambda \{ t_j \mid  H(t_j) \}  \hbox{exp} \left[ -\int_{0}^{\infty} Y(t) \lambda\{t \mid  H(t)\}dt  \right] \label{lld3}.
\end{eqnarray*}

If the intensity function is specified in terms of a parameter $\theta$, using the counting process notation, the log-likelihood can be expressed as 
\begin{eqnarray*}
    l(\theta)=\int_{0}^{\infty} Y(t)\hbox{log} \lambda\{t \mid  H(t);\theta\}dN(t) - \int_{0}^{\infty} Y(t) \lambda\{t \mid  H(t);\theta\}dt .
\end{eqnarray*}

\section{Estimating Equations of the Anderson-Gill Model}\label{sec:appen:ag}

Treating $d\mu_{0}(t)=\rho_{0}(t)dt$ as a parameter, we can have the estimating equation for $d\mu_{0}(t)$:
\begin{eqnarray*}
    \sum_{i=1}^{m} Y_{i}(t) \{ dN_{i}(t) - d\mu(t)  \} 
    = \sum_{i=1}^{m} Y_{i}(t) \{ dN_{i}(t) - \hbox{exp}(X^{T}_{i} \beta)d\mu_{0}(t)  \} = 0 .\label{esteqnlwyy}
\end{eqnarray*}

Then we obtain the profile likelihood estimates of $d\mu_{0}(t)$:
\begin{eqnarray*}
    d\widehat \mu_{0}(t; \beta) = \frac{\sum_{i=1}^{m} Y_{i}(t) dN_{i}(t)}{\sum_{i=1}^{m} Y_{i}(t) \hbox{exp}(X^{T}_{i} \beta)}. 
\end{eqnarray*}

Then, we have:
\begin{eqnarray*}
    U^{(A)}_{\beta}(\theta) 
    &=& \sum_{i=1}^{m}\int_{0}^{\tau} Y_{i}(t)X_{i}\{ dN_{i}(t) - \hbox{exp}(X^{T}_{i} \beta) d\mu_{0}(t) \} \\
    &=& \sum_{i=1}^{m}\int_{0}^{\tau} Y_{i}(t)X_{i} dN_{i}(t) - \sum_{i=1}^{m}\int_{0}^{\tau} Y_{i}(t)X_{i} \hbox{exp}(X^{T}_{i} \beta) \frac{\sum_{i=1}^{m} Y_{i}(t) dN_{i}(t)}{\sum_{i=1}^{m} Y_{i}(t) \hbox{exp}(X^{T}_{i} \beta)} \\
    &=& \int_{0}^{\tau} \sum_{i=1}^{m} Y_{i}(t) \left\{X_{i} -  \frac{\sum_{i=1}^{m} Y_{i}(t)X_{i} \hbox{exp}(X^{T}_{i} \beta)}{\sum_{i=1}^{m} Y_{i}(t) \hbox{exp}(X^{T}_{i} \beta)}  \right\} dN_{i}(t)  \\
    &=& \sum_{i=1}^{m}  \int_{0}^{\tau} Y_{i}(t) \left\{X_{i} -  \frac{\sum_{i=1}^{m} Y_{i}(t)X_{i} \hbox{exp}(X^{T}_{i} \beta)}{\sum_{i=1}^{m} Y_{i}(t) \hbox{exp}(X^{T}_{i} \beta)}  \right\} dN_{i}(t). 
\end{eqnarray*}

\section{Derivation of Estimating Equations for the NB Model with a Constant Baseline}\label{sec:appen:nb1}

By assuming $\lambda\{t \mid  H(t), \gamma\}=\gamma  \rho(t)$, the likelihood $L^{NB}_{i}(\theta, \phi)$ of each subject $i$ is
\begin{eqnarray*}
    L^{NB}_{i}(\theta, \phi) 
    &=& \int_{0}^{\infty} \prod_{j=1}^{n_i} \lambda \{ t_j \mid  H(t_j), \gamma_i \}  \hbox{exp} \left[ -\int_{0}^{\infty} Y(t) \lambda\{t \mid  H(t), \gamma_i \}dt  \right] f(\gamma_i) d\gamma_i  \\
    &=& \int_{0}^{\infty} \left[ \prod_{j=1}^{n_i} \gamma_{i} \rho_{i}(t_{ij})  \hbox{exp}\left\{ -\int_{0}^{\infty} Y_{i}(t) \gamma_{i} \rho_{i}(t) dt \right\} \right] \frac{1}{\Gamma(\phi^{-1}) \phi^{\frac{1}{\phi}} } \gamma_{i}^{\frac{1}{\phi}-1}\hbox{exp}\left(-\frac{\gamma_{i}}{\phi}\right)d\gamma_{i}  \\
    &=& \prod_{j=1}^{n_i} \rho_{i}(t_{ij}) \frac{1}{\Gamma(\phi^{-1}) \phi^{\frac{1}{\phi}} } \int_{0}^{\infty} \gamma_{i}^{n_i + \frac{1}{\phi}-1} \hbox{exp} \left[ -\gamma_i \left\{ \frac{1}{\phi}+\mu_{i}(\tau_i)  \right\}  \right] d\gamma_{i} \\
    &=& \prod_{j=1}^{n_i} \rho_{i}(t_{ij}) \frac{1}{\Gamma(\phi^{-1}) \phi^{\frac{1}{\phi}} }  \frac{\Gamma(\frac{1}{\phi} + n_i)}{\left\{ \frac{1}{\phi}+\mu_{i}(\tau_i)  \right\}^{\frac{1}{\phi}+n_i}} \\
    &=& \prod_{j=1}^{n_i} \rho_{i}(t_{ij}) \frac{1}{\Gamma(\phi^{-1}) \phi^{\frac{1}{\phi}} }  \frac{\Gamma(\frac{1}{\phi} + n_i)}{\left\{ \frac{1}{\phi}+\mu_{i}(\tau_i)  \right\}^{\frac{1}{\phi}+n_i}} \frac{\phi^{n_i}}{\phi^{n_i}} \frac{\mu_{i}(\tau_i)^{n_i}}{ \prod_{j=1}^{n_i} \mu_{i}(\tau_i) } \\
    &=& \left\{ \prod_{j=1}^{n_i} \frac{\rho_{0}(t_{ij})}{\mu_{0}(t_{ij})}  \right\} \frac{\Gamma(\frac{1}{\phi} + n_i)}{\Gamma(\frac{1}{\phi} )} \frac{\left\{\phi \mu_{i}(\tau_i)\right\}^{n_i}}{\{1+\phi \mu_{i}(\tau_i)\}^{\frac{1}{\phi} + n_i} }.
\end{eqnarray*}

The log-likelihood including all $m$ subjects is 
\begin{eqnarray*}
    l^{NB}(\theta, \phi) &=& \sum_{i=1}^{m} \Bigg[  \sum_{j=1}^{n_i} \{ \hbox{log}\rho_{0}(t_{ij}) - \hbox{log}\mu_{0}(\tau_{i})\} + n_i \hbox{log}\mu_{i}(\tau_i) -(n_i +\phi^{-1})  \hbox{log}\{1+ \phi\mu_{i}(\tau_i) \} \nonumber \\ 
    &+& n_i \hbox{log} \phi + \hbox{log} \{ (\phi^{-1} + n_i -1)  (\phi^{-1} + n_i -2)  \dots  \phi^{-1}    \} \Bigg] \\
    &=& \sum_{i=1}^{m} \Bigg[  \sum_{j=1}^{n_i}  \hbox{log}\frac{\rho_{0}(t_{ij})}{\mu_{0}(\tau_{i})} + n_i \hbox{log}\mu_{i}(\tau_i) -(n_i +\phi^{-1}) \hbox{log}\{1+ \phi\mu_{i}(\tau_i) \}  + \sum_{j=0}^{n_{i}^{*}} \hbox{log}(1+\phi j)  \Bigg]    
\end{eqnarray*}
where $n_{i}^{*}=\hbox{max}\{0, n_i-1\}$.

Then the estimating equations for $\phi$, $\alpha$, and $\beta$ based on the likelihood are
\begin{eqnarray*}
    U^{NB}_{\phi}(\theta, \phi) &=& \frac{\partial l^{NB}(\theta, \phi)}{\partial \phi} = \sum_{i=1}^{m} \left[ \phi^{-2}\hbox{log}\left\{ 1+\phi\mu_{i}(\tau_i) \right\} - (n_i+\phi^{-1}) \frac{\mu_{i}(\tau_i)}{1+\phi \mu_{i}(\tau_i)} + \sum_{j=0}^{n_{i}^{*}} \frac{j}{1+\phi j} \right], \nonumber \\
    U^{NB}_{\alpha}(\theta, \phi) &=& \frac{\partial l^{NB}(\theta, \phi)}{\partial \alpha} = \sum_{i=1}^{m} \left[  \left\{ \sum_{j=1}^{n_i} \frac{\partial \rho_{0}(t_{ij}) / \partial \alpha  }{\rho_{0}(t_{ij})}  \right\} - \frac{(1+ \phi n_i ) \hbox{exp}(X^{T}_{i} \beta ) }{1+ \phi\mu_{i}(\tau_i)  } \frac{\partial \mu_{0}(\tau_i)}{\partial \alpha}  \right], \nonumber \\
    U^{NB}_{\beta}(\theta, \phi) &=& \frac{\partial l^{NB}(\theta, \phi)}{\partial \beta} = \sum_{i=1}^{m} \frac{n_i - \mu_{i}(\tau_i) }{1+ \phi \mu_{i}(\tau_i) }  X_i.
\end{eqnarray*}

\section{Derivation of the Pseudo-Likelihood for the NB Model with an Unspecified Baseline}\label{sec:appen:nb2}

Let the event history $H^{*}_i(t)$ involve the total number of events up to but not including $t$, the corresponding event times, and the baseline covariates $\{N_{i}(t^{-}), t_{i1}, \dots, t_{iN_{i}(t^{-})}, X_i \}$. $H^{*}_i(t)$ differs from $H_i(t)$ in that it includes only covariates measured before treatment assignment, whereas $H_i(t)$ is defined to incorporate both baseline covariates and time-varying covariates collected after treatment assignment.

The P.D.F of $\gamma_i$ conditional on $\{N_{i}(t^{-}), t_{i1}, \dots, t_{iN_{i}(t^{-})} \}$ is 
\begin{eqnarray*}
    &&f\left\{\gamma_i \mid  N_{i}(t^{-}), t_{i1}, \dots, t_{iN_{i}(t^{-})}\right\} \\
    &&= \frac{\left\{ \prod_{j=1}^{N_{i}(t^{-})} \frac{\rho_{0}(t_{ij})}{\mu_{0}(t_{ij})}  \right\} \{ \gamma_i \mu_{i}(t) \}^{N_{i}(t^{-})} \hbox{exp}\{ -\gamma_i \mu_{i}(t) \} \frac{1}{\Gamma(\phi^{-1}) \phi^{\frac{1}{\phi}} } \gamma_{i}^{\frac{1}{\phi}-1}\hbox{exp}(-\frac{\gamma_{i}}{\phi})d\gamma_{i} }{\left\{ \prod_{j=1}^{N_{i}(t^{-})} \frac{\rho_{0}(t_{ij})}{\mu_{0}(t_{ij})}  \right\} \frac{\Gamma\{\frac{1}{\phi} + N_{i}(t^{-})\}}{\Gamma(\frac{1}{\phi} )} \frac{\{\phi \mu_{i}(t)\}^{N_{i}(t^{-})}}{\{1+\phi \mu_{i}(t)\}^{\frac{1}{\phi} + N_{i}(t^{-})} }} \\
    &&= \frac{1}{\Gamma\{N_{i}(t^{-}) +\phi^{-1}\} } \left\{  \frac{1+\phi \mu_{i}(t) }{\phi}\right\}^{N_{i}(t^{-}) +\phi^{-1}} \gamma_{i}^{N_{i}(t^{-}) +\phi^{-1}-1} \hbox{exp}\left\{ 
 -\gamma_i \frac{1+\phi \mu_{i}(t) }{\phi} \right\}
\end{eqnarray*}
which is the P.D.F of a gamma distribution with $\hbox{E}\{\gamma_i \mid  N_{i}(t^{-}), t_{i1}, \dots, t_{iN_{i}(t^{-})}\}=\frac{1+\phi N_{i}(t^{-})}{1+ \phi \mu_{i}(t)}$. 

Then the intensity function is:
\begin{eqnarray*}
    \lambda_{i}\{t \mid  H^{*}_i(t) \} &=& \int \lambda_{i}\{t \mid  H^{*}_i(t),\gamma_i \}  f\{\gamma_i \mid  H^{*}_i(t)\} d\gamma_i = \int  \gamma_i \rho_{i}(t)   f\{\gamma_i \mid  H^{*}_i(t)\} d\gamma_i \nonumber \\
    &=& \rho_{i}(t)  \hbox{E}\{\gamma_i \mid  H^{*}_i(t) \}   = \left\{ \frac{1+ \phi N_{i}(t^{-}) }{1+ \phi \mu_{i}(t) }  \right\} \rho_{i}(t)  = \left\{ \frac{1+ \phi N_{i}(t^{-}) }{1+ \phi \mu_{i}(t) }  \right\} \rho_{0}(t) \hbox{exp}(X^{T}_{i} \beta).
\end{eqnarray*}

If we adopt the semiparametric estimation procedure for $\rho_{0}(t)$, this is approximated by $\hat \rho_{0}(t)=\frac{\sum_{i=1}^{m} Y_{i}(t)  dN_{i}(t)}{\sum_{i=1}^{m} Y_{i}(t)  \hbox{exp}(X^{T}_{i} \beta)}$.

A pseudo-log-likelihood can be obtained by:
\begin{eqnarray*}
    l_{NB}(\beta;\phi)= \sum_{i=1}^{m} \int_{0}^{\infty} Y(t)\hbox{log} \lambda_i\{t \mid  H^{*}_i(t);\beta,\phi\}dN(t) - \int_{0}^{\infty} Y(t) \lambda_i\{t \mid  H^{*}_i(t);\beta,\phi\}dt .
\end{eqnarray*}

\bibliography{supplementary}